\documentclass[aps,prb,reprint,floatfix,superscriptaddress]{revtex4-2}

\usepackage[english]{babel} 
\usepackage{hyperref}
\usepackage{amssymb,amsmath}
\usepackage{graphicx}
\usepackage{float}
\usepackage{url}
\usepackage{multirow}
\usepackage{chemformula}
\usepackage{tabularx}
\usepackage{placeins}
\usepackage{mathtools}

\newlength{\smallpic}
\begin{document}

\title{Beyond the random phase approximation for calculating Curie temperatures in ferromagnets: application to Fe, Ni, Co and monolayer CrI$_3$}

\author{Varun Rajeev Pavizhakumari} 
\affiliation{CAMD, Computational Atomic-Scale Materials Design, Department of Physics, Technical University of Denmark, 2800 Kgs. Lyngby Denmark}
\author{Thorbjørn Skovhus} 
\affiliation{CAMD, Computational Atomic-Scale Materials Design, Department of Physics, Technical University of Denmark, 2800 Kgs. Lyngby Denmark}
\author{Thomas Olsen} 
\email{tolsen@fysik.dtu.dk}
\affiliation{CAMD, Computational Atomic-Scale Materials Design, Department of Physics, Technical University of Denmark, 2800 Kgs. Lyngby Denmark}

\begin{abstract}
The magnetic properties of solids are typically analyzed in terms of Heisenberg models where the electronic structure is approximated by interacting localized spins. However, even in such models the evaluation of thermodynamic properties constitutes a major challenge and is usually handled by a mean field decoupling scheme. The random phase approximation (RPA) comprises a common approach and is often applied to evaluate critical temperatures although it is well known that the method is only accurate well \textit{below} the critical temperature. In the present work we compare the performance of the RPA with a different decoupling scheme proposed by Callen as well as the mean field decoupling of interacting Holstein-Primakoff (HP) magnons. We consider three-dimensional (3D) as well as two-dimensional (2D) model systems where the Curie temperature is governed by anisotropy. In 3D, the Callen method is the most accurate in the classical limit, and we show that the Callen decoupling produces the best agreement with experiments for bcc Fe, fcc Ni and fcc Co with exchange interactions obtained from first principles. In contrast, for low spin systems where a  quantum mechanical treatment is pertinent, the HP and RPA methods are superior to the Callen decoupling. In 2D systems with magnetic order driven by single-ion anisotropy, it is shown that HP fails rather dramatically and both RPA and Callen approaches severely overestimates Curie temperatures. The most accurate approach is then constructed by combining RPA with the Callen decoupling of single-ion anisotropy, which yields the correct lack of order for $S=1/2$. We exemplify this by the case of monolayer CrI$_3$ using exchange constant extracted from experiments.
\end{abstract}

\maketitle

\section{Introduction}
The phase transition associated with loss of magnetic order constitutes a defining property of any magnet and the critical temperature where it happens is thus a crucial descriptor for magnetic materials. Besides being of fundamental interest, the critical temperature typically determines whether or not a given material may be of technological relevance. For example, a prerequisite for developing new materials for either spintronics, sensors or generators is a critical temperature that lies well above room temperature. Nevertheless, the theoretical prediction of critical temperature still poses a major challenge and the reliability of calculations is often difficult to validate. In particular, first principles simulations are typically not able to include finite temperature effects and the analysis of thermodynamic properties are commonly carried out by mapping the electronic structure to a Heisenberg model \cite{Illas1998,Xiang2013, Torelli2019b,Torelli2020}. The problem is then reduced to solving a Hamiltonian of localized spins coupled through exchange interactions. For the case of $S=1/2$ the one-dimensional Heisenberg model can be  diagonalized by the Bethe ansatz \cite{Karabach1997,Karbach1998} and for higher values of the spin, it may be treated accurately by the density matrix renormalization group \cite{PhysRevB.48.3844}. In two or three dimensions it is more difficult to apply accurate numerical techniques to the Heisenberg model. It is straightforward to perform \textit{classical} Monte Carlo simulations of the model and this approach seems to be gaining popularity in the literature \cite{Akturk2017,Torelli_2019, Kim2021,Li2022,Shen2022}. It is, however, far from clear that the Curie temperatures obtained from the classical model comprise reliable estimates. On the other hand, quantum Monte Carlo simulations \cite{Xiang1998} are computationally demanding and it is not obvious that the method is applicable to complex Heisenberg models with long-range exchange interactions obtained from first principles simulations using present day computational resources. Finally, there are several approximate mean field schemes that allow one to calculate thermodynamic properties of the Heisenberg model, but the accuracy of these are often difficult to assess. Such an assessment will be the main focus of the present work and we will show that commonly applied approximations are rather inaccurate and may lead to wrong validation of magnetic models due to spurious agreement of calculated Curie temperatures with experiments.

In 1940, Holstein and Primakoff (HP) \cite{holsteinprimakoff1940} showed that the spectrum of excitations that reduce the ground state magnetization by one Bohr magneton can be obtained by a bosonic transformation of the spin operators in the Heisenberg model. These excitations are known as magnons and yield a dispersion relation identical to the one derived by Bloch using a semi-classical analysis 10 years earlier \cite{Bloch1930}. The HP magnons are derived by truncating a series expansion of bosonic operators. This corresponds to assuming non-interacting magnons, whereas higher order terms correspond to magnon interactions. At elevated temperatures, the non-interacting dispersion relation becomes questionable since the magnetic properties are expected to be strongly influenced by the interactions between thermally excited magnons. In the HP approach, thermal effects can be included through a mean field decoupling of two-magnon interaction terms in the Heisenberg Hamiltonian \cite{Yosida1996}. This yields a thermally renormalized magnon dispersion, which tends to soften as the temperature rises and the critical temperature can be estimated as the point where the thermal occupation of magnons equals the ground state magnetization. While this approach is physically appealing, it is not \textit{a priori} clear if it is reliable to neglect higher order interactions or if the mean field decoupling is meaningful when the critical temperature is approached.

Another approach for treating thermal effects in the Heisenberg model was introduced by Bogoliubov and Tyablikov in 1958 \cite{tyablikov1959}. They considered the dynamic transverse susceptibility of the $S=1/2$ model and applied the random phase approximation (RPA) to perform a thermal decoupling of the higher order Green's functions that enters the equation of motion. This results in a thermally renormalized susceptibility with poles that may be interpreted as temperature-dependent energies of non-interacting quasi-particles. The method was generalized to arbitrary spin by Tahir-Keli and ter Haar \cite{tahir1962} and later simplified by Callen \cite{callen1963}. The simplicity of that approach is appealing and gives rise to an analytical expression for the critical temperature, which has been widely applied in the literature. The decoupling scheme does, however, assume that spin fluctuations along the magnetization direction may be ignored, which becomes dubious when the critical temperature is approached. Moreover, the RPA magnon dispersion may be seen as an approximation to the HP magnon dispersion \cite{Yosida1996}, which is thus regarded as more accurate \cite{Jensen1991} than RPA. We finally note that the HP decoupling scheme is sometimes referred to as RPA as well \cite{Yosida1996, Torelli_2019}, but in the present work we will reserve the term for the Tyablikov Green's function approach.

In 1963 Callen proposed a different decoupling scheme for the Green's function where the mean field itself depends on the magnetization \cite{callen1963}. On physical grounds this is expected to be much more accurate at high temperatures, whereas the approach reduces to the RPA method at low temperatures. The physical foundation and numerical results of the Callen decoupling appear superior to that of RPA, but the method seems to be overlooked in the literature. This is likely due the simplicity of RPA although it is rather straightforward to solve the self-consistent equations that will yield the Curie temperature from the Callen decoupling.

The RPA has been applied rather extensively to evaluate critical temperatures for transition metals \cite{Kudrnovsky2001} and has been used to benchmark exchange interactions evaluated by first principles by comparing calculated critical temperatures to experimental values \cite{Bruno2003, Katsnelson2004}. The most common approach for extracting Heisenberg exchange constants are by means of density functional theory (DFT) combined with the magnetic force theorem (MFT) \cite{LIECHTENSTEIN198765}. This approach only yields accurate magnon dispersion in the long wavelength limit, but it has been shown by Bruno that a better approximation for an adiabatic dispersion can be obtained by a simple renormalization of the MFT exchange constants \cite{Bruno2003}. The DFT-MFT magnon dispersion for fcc Ni is in good agreement with the experimental one but the associated RPA Curie temperature is severely underestimated \cite{Katsnelson2004}. On the other hand, the renormalized dispersion appears to be in disagreement with the experimental dispersion, but the associated RPA Curie temperature agrees well with the experimental result \cite{Bruno2003}. This discrepancy has given rise to some controversy regarding the "correct way" of calculating exchange constants, but the arguments seem to assume that the RPA results are valid once the exchange constants are known, which is far from obvious. Similarly, due to a booming interest in two-dimensional (2D) magnetism \cite{Huang2017, Gibertini2019}, the exchange constants in monolayer CrI$_3$ have been evaluated in a large number of recent works \cite{Olsen2017,Lado2017,Xu2018a,Torelli_2019,Olsen2019,Lu2019,Pizzochero2020,Shen2022} and the comparison of the experimental Curie temperature with those obtained from classical Monte Carlo simulations has been used to benchmark the performance of various functionals regarding the accuracy of magnetic properties \cite{Torelli_2019,Lu2019}. Again it is expected, that such an approach may lead to wrong conclusions due to the errors associated with the calculation of critical temperatures rather than the underlying DFT calculations.

A prominent difference between 2D and three-dimensional (3D) magnetism is that magnetic order in 2D cannot be driven by spontaneous symmetry breaking \cite{Mermin1966}, but requires explicit symmetry breaking by magnetic anisotropy. The single-ion anisotropy energy is the simplest way to include magnetic anisotropy in the Heisenberg model \cite{Torelli_2019}, but the mean-field decoupling discussed in either of the methods mentioned above entails an ambiguity related to the ordering of operators prior to decoupling. Due to the increasing interest in 2D magnetism it is pertinent to resolve this ambiguity, which can have a strong influence on the model predictions. For example, single-ion anisotropy in 2D $S=1/2$ systems cannot induce magnetic order at finite temperatures and whether or not this result is obtained in RPA depends on an arbitrary choice of operator ordering prior to the mean field decoupling.

In the present work we perform a thorough comparison of the Callen method and compare with results obtained from RPA and HP decoupling as well as Weiss mean-field theory and classical Monte Carlo simulations. In particular, we benchmark the methods in the classical limit where the exact result can be obtained numerically from Monte Carlo (MC) simulations. The Callen and HP methods are found to yield excellent agreement with MC, whereas RPA underestimates the critical temperatures. The approach is applied to the cases of fcc Ni, bcc Fe and fcc Co with exchange parameters obtained from DFT and it is shown that the Callen approach yields the best agreement with experiments for Ni and Co whereas HP seems to be superior for bcc Fe. We then analyze 2D magnetic order driven by magnetic anisotropy and discuss the subtleties related to mean field decoupling of the single-ion anisotropy term. Finally, we discuss the case of monolayer CrI$_3$ where we fix the exchange constant from the measured optical magnon energies. The value disagrees with the previous claims of accurate exchange constant from DFT+U (a factor two larger than LDA results) and we argue that this is due to a severe underestimation of the critical temperature from classical MC simulations.

The paper is organized as follows. In Sec. \ref{sec:theory} we summarize the basic theory and equations that need to be solved in the HP, RPA and Callen methods. In Sec. \ref{sec:results} we present the results starting with a model lattice in 3D followed by the cases of bcc Fe, fcc Ni and fcc Co with exchange constant obtained from DFT. We then consider a 2D model lattice and discuss the effect of single-ion anisotropy using different decoupling schemes in each of the three approaches. Finally we calculate the critical temperature in monolayer CrI$_3$ with exchange constants obtained from experiments. In the appendix we provide the generalization of the HP, RPA and Callen methods for multiple magnetic sites in the unit cell.

\section{Theory}\label{sec:theory}
We start by considering the isotropic Heisenberg model for a magnetic Bravais lattice, which we represent as
\begin{equation}\label{heisenbrghamiltonian}
    H = -\frac{1}{2}\sum_{i\neq j} J_{ij}\, \mathbf{S}_{i} \cdot \mathbf{S}_{j},
\end{equation}
where $\mathbf{S}_{i}$ is the spin operator of site $i$ positioned at the lattice vector $\mathbf{R}_i$ and summation over both $i$ and $j$ indices is implied. The inclusion of anisotropy terms will be considered in section \ref{subsec:2Dmagnon}. For ferromagnets where the spins are aligned along some quantization axis in the ground state---here chosen as the $z$-axis---it is furthermore convenient to introduce the spin raising and lowering operators
\begin{equation}
S^\pm_i=S_i^x \pm iS_i^y,
\end{equation}
in the representation of which
\begin{equation}
    H = - \frac{1}{2} \sum_{i\neq j} J_{ij} \left(\frac{1}{2}\left[S_i^+ S_j^- + S_i^- S_j^+\right] + S_i^z S_j^z\right),
    \label{eq:hamiltonian in circular coordinates}
\end{equation}
with commutation relations (in atomic units)
\begin{subequations}\label{spin-comm}
\begin{align}
[S_i^z,S_j^\pm] &= \pm S_i^\pm\delta_{ij}, \label{spin-comm-1}\\
[S_i^+,S_j^-] &= 2S_i^z\delta_{ij}, \label{spin-comm-2}\\
[S_i^\pm,H] &= \pm\sum_{j\neq i}J_{ij}\Big(S_i^\pm S_j^z-S_i^zS_j^\pm\Big). \label{spin-comm-H}
\end{align}
\end{subequations}
It is then rather straightforward to diagonalize the one-magnon sector (excitations that lower the total spin by one) of $H$ using either the Holstein-Primakoff approach, semiclassical spin-wave theory or direct application of Bloch's theorem. However, for thermal properties one needs to include all possible excitations and these are not easily obtained from the model. Instead, one may apply various approximate schemes.

In the following we will summarize the basic results obtained from a thermal mean-field approximation, the Holstein-Primakoff approach as well as the Green's function based RPA and Callen decoupling schemes.

\subsection{Mean Field Theory}
The mean field approximation (MFA) is the simplest approach for obtaining an estimate for the Curie temperature. The spin operators are written as $\mathbf{S}_i=\langle\mathbf{S}\rangle+\delta\mathbf{S}_i$ where $\langle\mathbf{S}\rangle$ represents the thermal average of the spin operator. Neglecting terms quadratic in $\delta\mathbf{S}_i$ and assuming ferromagnetic order along $z$ (in which case $\langle S^\pm\rangle = 0$),
\begin{equation}\label{eq:H_mf}
    H^\mathrm{MFA} = - \frac{1}{2}\sum_i J_0 \left(2S^z_i - \langle S^z\rangle\right) \langle S^z\rangle
\end{equation}
where $J_0 = \sum_{i\neq 0} J_{0i}$. With each site decoupled, the energy per site is (up to a constant) given by
\begin{equation}
    E_i^\mathrm{MFA} = - J_0 S_i^z \langle S^z \rangle,
    \label{eq:mfa energy per site}
\end{equation}
and the thermal average of $S_i^z$ can be calculated as a sum over the eigenvalues $m_s$ to the $S^z$ operator,
\begin{equation}\label{mfa-mag}
    \langle S^z \rangle = \frac{1}{\mathcal{Z}_0}\sum_s m_s e^{J_0 m_s \langle S^z \rangle/k_\mathrm{B}T},
\end{equation}
where $\mathcal{Z}_0$ is the partition function for a single site, defined via Eq. \eqref{eq:mfa energy per site}. Finding the minimum temperature $T$ under which Eq. \eqref{mfa-mag} ceases to yield a finite solution for $\langle S^z\rangle$, the MFA Curie temperature is identified as \cite{Yosida1996}
\begin{equation}\label{eq:T_mf}
    k_\mathrm{B} T_\mathrm{c}^\mathrm{MFA} = \frac{J_0 S(S+1)}{3} ,
\end{equation}
where $S$ is the maximum eigenvalue of $S^z$.

The simplicity of the MFA is appealing and it may readily be generalized to spiral order (with antiferromagnets as a special case). The neglect of explicit correlations is, however, often too simple to produce reliable quantitative estimates of Curie temperatures as will be shown below. In particular, for 2D systems the prediction of finite Curie temperatures in the isotropic Heisenberg model is in direct contradiction with the Mermin-Wagner theorem \cite{Mermin1966} and the mean-field treatment becomes qualitatively wrong in that case.

\subsection{Holstein-Primakoff approach}
In the Holstein-Primakoff (HP) method \cite{holsteinprimakoff1940}, the spin operators are expressed in terms of bosonic operators satisfying $[a_i,a^\dag_j]=\delta_{ij}$. It may be shown that the commutator relations for the spin operators are conserved by the relations
\begin{subequations}
\begin{align}
        S_i^+ & = \sqrt{2S-a_i^\dagger a_i}\, a_i \label{hp_spin+},\\
        S_i^- & = a_i^\dagger \sqrt{2S-a_i^\dagger a_i} \label{hp_spin-},\\
        S_i^z & = S - a_i^\dagger a_i \label{hp_spinz},
\end{align}
\end{subequations}
where $S$ again denotes the maximum eigenvalue of $S^z$. Expanding the square roots in the Hamiltonian \eqref{eq:hamiltonian in circular coordinates} and retaining only terms up to quadratic order, the problem becomes exactly solvable \cite{Toth2015}, generating a series of eigenstates that are interpreted as magnons. Higher order terms represent magnon-magnon interactions, which become important at elevated temperatures. Thus, for any accurate treatment of thermal properties these need to included and one is left with an intractable interacting bosonic Hamiltonian. Nevertheless, if the expansion is truncated at fourth order in bosonic operators, the interaction terms can be handled in a Hartree-Fock type of mean-field decoupling. This is most conveniently done in Fourier space. Defining
\begin{equation}\label{hp-q-transform}
   a_{\mathbf{q}} = \frac{1}{\sqrt{N_q}}\sum_{\mathbf{q}} a_i e^{-i\mathbf{q}\cdot \mathbf{R}_{i}}
\end{equation}
results in fourth order terms like $a^\dag_\mathbf{q}a^\dag_\mathbf{q'}a_\mathbf{q''}a_\mathbf{q+q'-q''}$, which may be approximated by a set of terms of the form $a^\dag_\mathbf{q}a_\mathbf{q''}\langle a^\dag_\mathbf{q'}a_\mathbf{q+q'-q''}\rangle$, where $\langle\ldots\rangle$ denotes thermal averaging. Under the assumption that the thermal average vanishes unless bosons are created and annihilated at the same $\mathbf{q}$, this results in a mean-field Hamiltonian that is quadratic in bosonic operators and depends on the Bose factors $n_\mathbf{q}^\mathrm{B}(T)=\langle a^\dag_\mathbf{q}a_\mathbf{q}\rangle$ as well as the Fourier transform of the the exchange constants
\begin{equation}
    J_{\mathbf{q}} = \sum_{\mathbf{R}_i} J_{0i} e^{i{\mathbf{q}} \cdot \mathbf{R}_i} \label{Jij-q-space}.
\end{equation}
The Hamiltonian is diagonal in $\mathbf{q}$ and the thermally renormalized magnon dispersion is given by \cite{Yosida1996}
\begin{align}\label{HP_magnon_energy}
    \omega_{\mathbf{q}}^\mathrm{HP} &=  \langle S^z \rangle (J_0 - J_{\mathbf{q}})+ \frac{1}{N_q}\sum_{\mathbf{q}^\prime}(J_{{\mathbf{q}}^\prime} - J_{\mathbf{q} - {\mathbf{q}}^\prime})n^\mathrm{B}_{{\mathbf{q}}^\prime},
\end{align}
where the thermal average of $S^z$ (magnetization) may be expressed as
\begin{equation}\label{HP_mag}
    \langle S^z \rangle = S - \phi,
\end{equation}
with
\begin{equation}\label{phi}
    \phi = \frac{1}{N_q} \sum_{\mathbf{q}} n^\mathrm{B}_{\mathbf{q}}(T).
\end{equation}
Since the Bose factors in Eq. \eqref{HP_magnon_energy} has an implicit dependence on the magnon energies, Eqs. \eqref{HP_magnon_energy}-\eqref{phi} have to be solved self-consistently.

The magnon energy \eqref{HP_magnon_energy} reduces to the result from linear spin-wave theory or semi-classical Bloch theory if the temperature vanishes. At elevated temperatures the magnetization is reduced and the magnon dispersion softens as a result. Conversely, a softening of the magnon dispersion leads to a reduced magnetization. It is thus natural to identify the critical temperature as the point where the magnetization vanishes. However, the theory typically breaks down before this point due to the appearance of negative magnon energies, which renders the Bose factors ill defined. Nevertheless, since the derivative of the magnetization as a function of temperature approaches $-\infty$ at this point, it is natural to associate the corresponding temperature with the ferromagnetic phase transition, although the discontinuity of the magnetization indicates a somewhat pathological behaviour of the model close to the Curie temperature.

\subsection{Random phase approximation}
A different line of approach for including thermal effects on the magnetic order was developed by Bogliubov and Tyablikov \cite{tyablikov1959} and is widely referred to as the Random Phase Approximation (RPA) in the literature. It is similar in spirit to the decoupling of fourth order terms in the Holstein-Primakoff method (which is also sometimes referred to as RPA \cite{Yosida1996}), but is based on a Green's function treatment. The basic quantity of interest is the retarded Green's function
\begin{equation}\label{arbGfn-def}
    G_{ij}^{+-}(t) = -i \theta(t)\langle [S_i^+(t), S_j^-] \rangle\equiv\langle\langle S_i^+(t); S_j^-\rangle\rangle,
\end{equation}
which comprises the dynamic transverse susceptibility of the system. The Fourier transform is represented as
\begin{align}\label{arbGfn-def_fourier}
    G_{ij}^{+-}(\omega) &= \int_{-\infty}^\infty dt\, e^{i(\omega+i\eta)t}\langle\langle S_i^+(t); S_j^- \rangle\rangle\notag\\&\equiv\langle\langle S_i^+; S_j^- \rangle\rangle_\omega,
\end{align}
where $\eta$ is an infinitesimal positive frequency. It is straightforward to show that $G_{ij}^{+-}(\omega)$ will have poles at the fundamental excitation energies of the Heisenberg model.
Fourier transforming the equation of motion for the Green's function and applying the commutation relations \eqref{spin-comm} yields
\begin{subequations}
\begin{align}\label{eqnofmotion-Gfn}
(\omega+i\eta)G_{ij}^{+-}(\omega) &= 
\langle [S^+_i, S^-_j] \rangle + \langle \langle [S_i^+ , H]; S_{j}^- \rangle \rangle_\omega \\
&= 2\langle S^z\rangle \delta_{ij} \nonumber\\
&+ \sum_{k\neq i} J_{ik}\langle \langle S_i^+ S^z_k - S^z_i S_k^+; S_{j}^- \rangle \rangle_\omega.
\label{eq:eom after comm}
\end{align}
\end{subequations}
The second term of Eq. \eqref{eqnofmotion-Gfn} can be expressed in terms of higher order Green's functions that need to be approximated in order to obtain a closed expression for $G^{+-}_{ij}$.
In the RPA, it is assumed that fluctuations in $S_i^z$ are small and it is replaced by its average value $\langle S^z\rangle$ (which is independent of the site index for ferromagnets with equivalent sites).
The central approximation is thus
\begin{align} \label{rpa-approx}
    \langle \langle [S_i^z S_j^+](t); S^-_k\rangle \rangle &\approx \langle S^z \rangle \langle \langle S_j^+(t); S^-_k \rangle \rangle, \quad i\neq j\notag\\  
    &=\langle S^z \rangle G_{jk}^{+-}(t).
\end{align}
This is expected to be true at low temperatures, but is probably not well justified when approaching the critical temperature. Introducing the Fourier transform
\begin{align}
    G_{\mathbf{q}}^{+-}(\omega)= \sum_{i} G_{0i}^{+-}(\omega) e^{i{\mathbf{q}} \cdot \mathbf{R}_i} \label{Gfn-q-space},
\end{align}
the equation of motion can be solved yielding
\begin{equation}\label{G_rpa}
    G_{\mathbf{q}}^{+-}(\omega)=\frac{2\langle S^z\rangle}{\omega-\omega^\mathrm{RPA}_\mathbf{q}+i\eta},
\end{equation}
with
\begin{equation}\label{rpa-magnons3d}
    \omega^\mathrm{RPA}_{\mathbf{q}} = \langle S^z\rangle (J_0  - J_{\mathbf{q}} ).
\end{equation}
The poles $\omega^\mathrm{RPA}_{\mathbf{q}}$ are identified as the temperature-dependent magnon energies and reduces to the result of linear spin-wave theory at $T=0$. At finite temperatures, they are renormalized by the magnetization in a fashion identical to the first term in the HP dispersion \eqref{HP_magnon_energy}. It is thus tempting to regard this as an approximation to the Holstein-Primakoff scheme with mean-field treatment of interacting bosons, but this picture is not completely valid. The reason is that there is no {\it a priori} reason to regard the excitations as proper bosons in RPA and the magnetization $\langle S^z\rangle$ is not simply obtained from the Bose distribution associated with $\omega^\mathrm{RPA}_{\mathbf{q}}$, but has to be evaluated from the Green's function itself.

For the case of $S=1/2$, it is straightforward to obtain an expression for the magnetization since we then have $\langle S^z\rangle=1/2-\langle S^-S^+\rangle=1/2-2\langle S^z\rangle\phi$ where the second equality is obtained from the fluctuation-dissipation theorem and $\phi$ was defined in Eq. \eqref{phi}. As a result, $\langle S^z\rangle=1/(2+4\phi)$, and it is only in the low temperature limit $\phi\ll1$ where $\langle S^z\rangle=1/2-\phi$ and an interpretation in terms of bosonic excitations can be established. For $S>1/2$, it is a bit more involved to obtain an expression for the magnetization, but it may be accomplished by the introduction of auxiliary Green's functions \cite{tahir1962,callen1963}. The result is
\begin{equation}\label{mag}
    \langle S^z \rangle = \frac{(S-\phi)(1+\phi)^{2S+1} + (S+1+\phi)\phi^{2S+1}}{(1+\phi)^{2S+1} - \phi^{2S+1}},
\end{equation}
which again reduces to $\langle S^z\rangle=S-\phi$ at low temperatures. Eqs. \eqref{rpa-magnons3d} and \eqref{mag} thus comprise a closed set of equations that needs to be solved self-consistently in order to determine the magnetization at a given temperature. The critical temperature can be determined as the point where the magnetization vanishes and in contrast to the Holstein-Primakoff method it decreases continuously towards zero as the temperature approaches the Curie temperature from below.

It is possible to obtain a closed expression for the Curie temperature by performing an expansion of Eq. \eqref{mag} in the limit of large $\phi$ (as expected near the Curie temperature) and Taylor expanding the Bose factors in terms of $\omega^\mathrm{RPA}_\mathbf{q}$ (since these are proportional to $\langle S^z\rangle$). In general this leads to 
\begin{equation}\label{Tc3d-gen}
    k_\mathrm{B} T_\mathrm{c} = \lim_{\langle S^z \rangle \rightarrow 0} \frac{S(S+1)}{3} \biggl( \frac{1}{N_q} \sum_{\mathbf{q}} \frac{\langle S^z \rangle}{\omega_{\mathbf{q}}(\langle S^z \rangle)}\biggr )^{-1},
\end{equation}
and for the RPA dispersion \eqref{rpa-magnons3d} one obtains
\begin{equation}\label{Tc3d-rpa}
    k_\mathrm{B} T_\mathrm{c}^\mathrm{RPA} =\frac{S(S+1)}{3} \biggl( \frac{1}{N_q} \sum_{\mathbf{q}} \frac{1}{J_0-J_\mathbf{q}}\biggr )^{-1}.
\end{equation}
Similarly, if one applies the mean-field Hamiltonian \eqref{eq:H_mf} in Eq. \eqref{eqnofmotion-Gfn} one obtains a "mean-field dispersion" of $\omega^\mathrm{MFA}=\langle S^z\rangle J_0$, which reproduces \eqref{eq:T_mf} when inserted into \eqref{Tc3d-gen}.

\subsection{Callen decoupling}
In essence the RPA replaces $S^z$ with $\langle S^z\rangle$ and thus neglects fluctuations in $S^z$. While this is expected to be a good approximation at low temperatures it is likely to break down when the critical temperature is approached. To improve upon this, an alternative decoupling scheme was proposed by Callen \cite{callen1963}. The idea is based on the fact that different representations of the $S^z$ operator will lead to different mean field approximations for the Green's function. Specifically, we may write $S^z$ in three distinct ways:
\begin{align}
    S_i^z &= S(S+1) - S_i^-S_i^+ - {S_i^z}^2 \label{Sz-cd1},\\
    S_i^z &= -S(S+1) + S_i^+S_i^- + {S_i^z}^2 \label{Sz-cd2},\\
    S_i^z &= \frac{S_i^+S_i^- - S_i^-S_i^+}{2} \label{Sz-cd3}.
\end{align}
Introducing a parameter $\alpha\in\{-1,0,1\}$, Eqs. \eqref{Sz-cd1}-\eqref{Sz-cd3} can be combined into a single expression:
\begin{equation}\label{Sz-cd}
    S_i^z = \alpha (S(S+1)- {S_i^z}^2) + \frac{1}{2}(1-\alpha)S_i^+S_i^- - \frac{1}{2}(1+\alpha)S_i^-S_i^+.
\end{equation}
Inserting this into the higher order Green's functions appearing in the equation of motion \eqref{eq:eom after comm} and symmetrically decoupling the $S^-S^+$ terms, Callen arrives at a unified decoupling \cite{callen1963}
\begin{align}\label{cd-approx}
    \langle \langle S_i^z S_j^+;S^-_k \rangle \rangle \approx \langle S^z \rangle &\langle \langle S_j^+;S^-_k \rangle \rangle \\ & - \alpha \langle S_i^- S_j^+ \rangle \langle \langle S_i^+; S^-_k\rangle \rangle,\quad i\neq j\notag.
\end{align}
Now, instead of fixing the choice of $\alpha$, Callen pointed out that $\alpha\in[-1,1]$ can be chosen to vary continuously with temperature, thus letting the type of correlations to neglect be temperature dependent.  
Based on a view that Eq. \eqref{Sz-cd3} best represents the deviations of $S^z_i$ from 0, Callen argued that $\alpha\rightarrow 0$ for $\langle S^z\rangle\rightarrow 0$, and in order to make fluctuations in $S^z_i$ on the order of 1 in the low temperature limit, he arrived at the functional form $\alpha=\frac{\langle S^z \rangle}{2S^2}$.
Solving the equation  of motion in the Callen decoupling (CD) then gives
\begin{equation}\label{G_cd}
    G_{\mathbf{q}}^{+-}(\omega)=\frac{2\langle S^z\rangle}{\omega-\omega^\mathrm{CD}_\mathbf{q}+i\eta}
\end{equation}
with
\begin{align}\label{cd-magnon3d}
    &\omega^\mathrm{CD}_{\mathbf{q}} = \langle S^z \rangle (J_0  - J_{\mathbf{q}} ) + \frac{\langle S^z\rangle^2}{N_qS^2} \sum_{\mathbf{q}^\prime} \bigl(J_{{\mathbf{q}}^\prime} - J_{\mathbf{q} - {\mathbf{q}}^\prime} \bigr) n^\mathrm{B}_{{\mathbf{q}}^\prime}.
\end{align}
Comparing with the RPA dispersion \eqref{rpa-magnons3d}, we note the appearance of additional terms arising from the second decoupling term in Eq. \eqref{cd-approx}. The Bose factor in this term introduces an additional temperature renormalization to the magnon energies similar to the HP dispersion \eqref{HP_magnon_energy}. In fact, the only difference between Eqs. \eqref{HP_magnon_energy} and \eqref{cd-magnon3d} is the prefactor $\langle S^z\rangle^2/S^2$, which is absent in HP dispersion. This factor, however, ensures 
that the additional renormalization of the RPA dispersion is decreased in weight
as the Curie temperature is approached from below---in contrast to the HP approach where the renormalization leads to negative magnon energies at a finite value of $\langle S^z\rangle$. Hence, in a certain sense the Callen decoupling scheme can be viewed as intermediate between the RPA and HP methods.

\subsection{Magnetic anisotropy}\label{subsec:2Dmagnon}
The inclusion of magnetic anisotropy in the Heisenberg model \eqref{heisenbrghamiltonian} introduces additional complications in the decoupling scheme. In particular, single-ion anisotropy terms give rise to an ambiguity that leads to results depending on the ordering of operators prior to decoupling. As we will see, the Holstein-Primakoff method implies a natural choice of ordering, which reproduces the correct magnon dispersion in the $T\rightarrow0$ limit, whereas the RPA decoupling only produces the correct dispersion using a particular (but in principle arbitrary) choice of operator ordering. In contrast, the Callen decoupling gives the correct dispersion regardless of the choice of ordering. For 2D materials, magnetic anisotropy is crucial for maintaining order at any finite temperature and an accurate treatment of anisotropy terms in the decoupling is critical.

We restrict ourselves to uniaxial anisotropy, which we represent by the additional terms in the Hamiltonian:
\begin{equation}\label{2dheisenberghamiltonian}
    \Delta H = - \frac{1}{2} \sum_{i\neq j} \lambda_{ij} S_i^z S_j^z - A \sum_i ({S_i^z})^2,
\end{equation}
where $\lambda_{ij}$ are anisotropic exchange constants and $A$ is the single-ion anisotropy constant. The anisotropy breaks the rotational symmetry of the model explicitly and leads to a ferromagnetic ground state polarized along $z$ if (as we will see) $S\sum_{i\neq0}\lambda_{0i}+A(2S-1)>0$.

The Holstein-Primakoff transformation of the additional terms \eqref{2dheisenberghamiltonian} only involves \eqref{hp_spinz} and for the anisotropic exchange (involving $\lambda_{ij}$), it is straightforward to decouple the bosonic operators in a manner similar to the isotropic terms. The single-ion anisotropy (involving $A$), however, gives rise to a fourth order term involving bosonic operators at the same site, which may be written in two equivalent ways:
\begin{align}\label{sia-reorder}
   a_i^\dagger a_i a_i^\dagger a_i = a_i^\dagger a_i^\dagger a_i a_i + a_i^\dagger a_i.
\end{align}
and applying the decoupling to the left hand side will differ from decoupling the right hand side. 
Nevertheless, only the fourth order term on the right hand side becomes irrelevant in the single-magnon sector of the spectrum and it is thus natural to apply the decoupling to this term and include the quadratic term on the right hand side in linear spinwave theory. Inclusion of the two terms in Eq. \eqref{2dheisenberghamiltonian}, thus yields the additional terms in the dispersion:
\begin{align}\label{hp_correction}
    \Delta\omega^\mathrm{HP}_\mathbf{q} =A(2S-1)+\lambda_0\langle S^z\rangle-\frac{1}{N_q}\sum_{\mathbf{q}'}(\lambda_\mathbf{q-q'}+4A)n^\mathrm{B}_{\mathbf{q}'},
\end{align}
with
\begin{equation}\label{lambda-q-space}
   \lambda_\mathbf{q} = \sum_{i\neq0} \lambda_{0i} e^{i{\mathbf{q}} \cdot\mathbf{R}_i}.
\end{equation}
At vanishing temperature only the first two terms in Eq. \eqref{hp_correction} survive and result in a fundamental spin-wave gap of $A(2S-1)+\lambda_0S$. It is also clear that the first term becomes irrelevant for $S=1/2$, which is expected since the single-ion anisotropy term in Eq. \eqref{2dheisenberghamiltonian} is then a constant that cannot influence the spectrum. The single-ion anisotropy does, however, influence the spectrum at finite temperatures (through the last term of Eq. \eqref{hp_correction}), which should be regarded as a fundamental flaw of the theory for $S=1/2$.

In the case of RPA, 
one is faced with the question of how to decouple the single-ion anisotropy
\begin{align}\label{standard_order}
-A\langle\langle[S^+_i,(S^z_i)^2];S^-_j\rangle\rangle=A\langle\langle S^z_iS^+_i+S^+_iS^z_i;S^-_j\rangle\rangle.
\end{align}
A straightforward replacement of $S^z_i$ with its average
results in additional terms in the dispersion of the form
\begin{align}\label{rpa_correction}
    \Delta\omega^\mathrm{RPA}_\mathbf{q} =A2\langle S^z\rangle+\lambda_0\langle S^z\rangle,
\end{align}
which are independent of $\mathbf{q}$. This is somewhat similar to the first terms of Eq. \eqref{hp_correction}, except that the factor of $2S-1$ has been replaced by $2\langle S^z\rangle$, thus failing to reproduce the irrelevance of single-ion anisotropy for $S=1/2$. 
However, using the commutator relation \eqref{spin-comm-1}, we may also base the decoupling on the expressions
\begin{align}
-A\langle\langle[S^+_i,(S^z_i)^2];S^-_j\rangle\rangle=&A\langle\langle 2S^z_iS^+_i-S^+_i;S^-_j\rangle\rangle\label{rpa_correct_ordering}\\
=&A\langle\langle 2S^+_iS^z_i+S^+_i;S^-_j\rangle\rangle,
\end{align}
which replaces the factor of $2\langle S^z\rangle$ in the first term of Eq. \eqref{rpa_correction} by $2\langle S^z\rangle\mp1$. 
Unlike the case of the Holstein-Primakoff decoupling there is no natural choice of which expression to use and this ambiguity (which may or may not lead to a correct physical results) renders the RPA decoupling for single-ion anisotropy rather dubious. Moreover, even 
if one chooses to base the decoupling on Eq. \eqref{rpa_correct_ordering}, (which leads to correct low-temperature dispersion for $S=1/2$), 
the resulting magnetization does not vanish smoothly as the temperature approaches the Curie temperature. Instead, a requirement of having a continuous magnetization as a function of temperature fixes the operator ordering to that of Eq. \eqref{standard_order} and the dispersion to $\eqref{rpa_correction}$.

Finally, the Callen decoupling scheme leads to additional terms in the dispersion given by
\begin{align}\label{callen_correction}
    \Delta\omega^\mathrm{CD}_\mathbf{q}=&A\Big(2\langle S^z\rangle-\frac{\langle S^z\rangle^2}{S^2}\Big)+\lambda_0\langle S^z\rangle\notag\\&-\frac{\langle S^z\rangle^2}{S^2N_q}\sum_{\mathbf{q}'}(\lambda_\mathbf{q-q'}+2A)n^\mathrm{B}_{\mathbf{q}'}.
\end{align}
This yields the correct irrelevance of single-ion anisotropy for $S=1/2$ at $T=0$, but acquires a dependence on $A$ for finite temperatures, albeit with a smaller prefactor (that decreases with $\langle S^z\rangle$) than in the Holstein-Primakoff method. Again, the dispersion relies on the choice of operator ordering and the present expression was obtained from \eqref{standard_order}. But in contrast to RPA, the CD leads to irrelevance of the single-ion anisotropy for $S=1/2$ at $T=0$ for any of the choices of operator ordering. However, the expression \eqref{standard_order} is the only one that leads to a continuous magnetization that vanishes smoothly as the Curie temperature is approached from below.

To summarize, the anisotropic exchange is straightforward to include in any of the three methods considered here, while the single-ion anisotropy terms lead to a somewhat ambiguous dependence on the operator ordering prior to the decoupling in the Green's function approaches. Nevertheless, if we require a magnetization that vanishes continuously as the critical temperature is approached, the choice of operator ordering is fixed and the Callen method is the only one that retains the irrelevance of single-ion anisotropy terms at low temperatures for $S=1/2$ systems. The operator ordering is less ambiguous in the Holstein-Primakoff decoupling, but that method always leads to discontinuous magnetization---even in the absence of anisotropy.

\section{Results}\label{sec:results}
We will start by benchmarking the different decoupling schemes for isotropic 3D systems in the classical limit. For simplicity we begin with a cubic lattice with nearest neighbor interactions, where the results may be compared to classical MC simulations. The methods are then applied to bcc Fe, fcc Ni and fcc Co with exchange constants obtained from first principles and the calculated critical temperatures are compared to experimental values. We then compare the methods for 2D ferromagnets, where anisotropy is decisive for obtaining finite Curie temperatures. We start by considering model Bravais lattices with nearest neighbor interactions and compare the classical limit to Monte Carlo simulations. Finally, we consider the case of CrI$_3$ where we obtain the Curie temperature based on Heisenberg parameters extracted from experimental magnon energies. It will be argued that classical Monte Carlo simulations lead to severe underestimation of the critical temperature and accurate critical temperatures obtained from such approaches in the literature seem to rely on significant overestimation of calculated exchange constants.

\subsection{Cubic lattice with nearest neighbor isotropic interactions}\label{sec:cubic}
\begin{figure}[tb]
    \centering
    \includegraphics[scale=0.3]{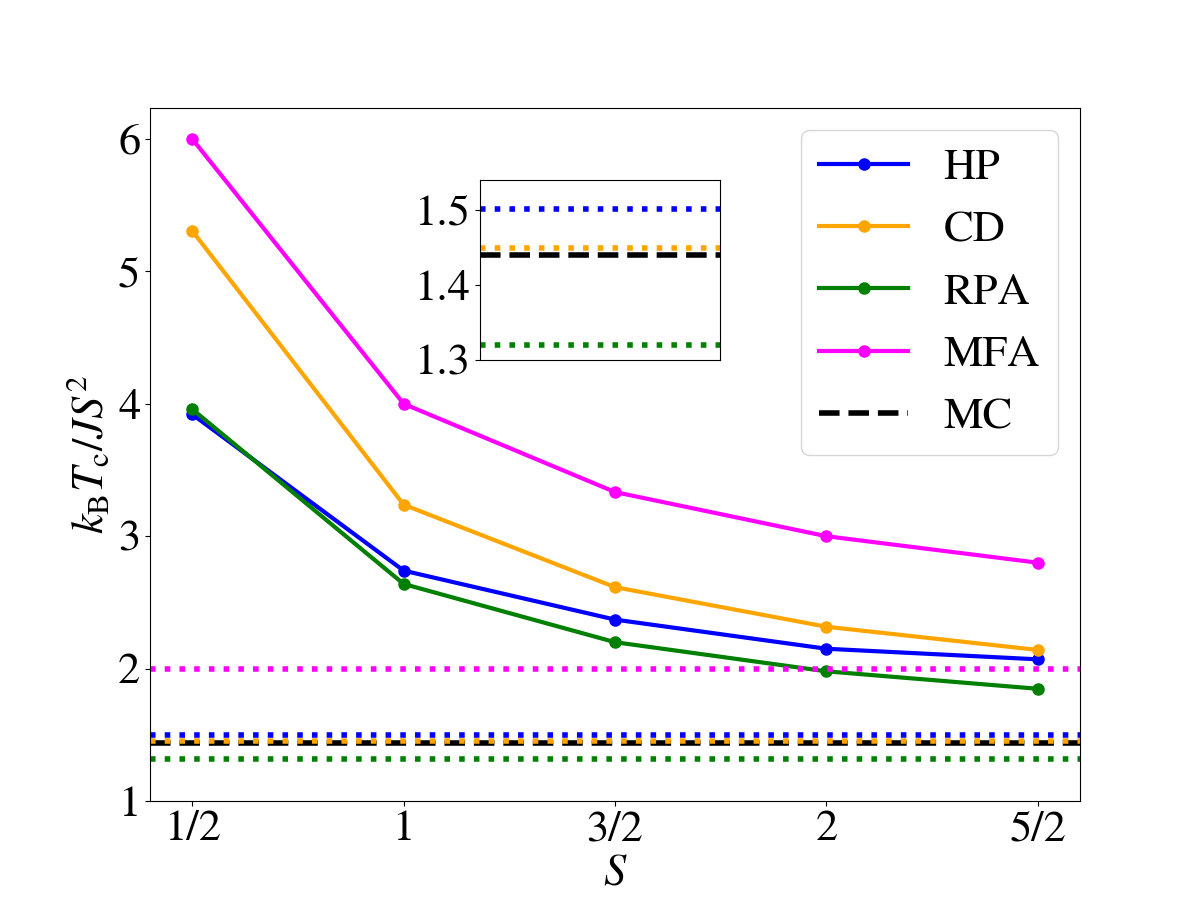}
    \caption{The variation of critical temperature with spin for the cubic lattice with nearest neighbor isotropic exchange. The horizontal lines are the classical limits for each of the methods.}
    \label{fig:s-classical3d}
\end{figure}
In Fig. \ref{fig:s-classical3d} we compare the Curie temperature obtained with MFA, RPA, HP, CD as well as classical Monte Carlo (MC) simulations for the simple cubic lattice with nearest neighbor interactions. All MC simulations are done by initializing a ferromagnetic configuration in a $20\times20\times20$ super cell. The critical temperature was obtained from the heat capacity which exhibits a peak at $T_\mathrm{C}$ and we checked that the results were converged with respect to super cell size and initial configuration of magnetic moments. The results are shown as a function of $S$ and the Curie temperature is shown in units of $JS^2/k_\mathrm{B}$ such that the classical result is independent of $S$. We note that the spin dependence of the MFA and RPA results are given exactly by $S(S+1)$ and a similar dependence is roughly observed for HP and CD. At low values of $S$, RPA and HP is in agreement but start deviating at higher spin where CD and RPA starts converging towards similar values. In the classical limit ($S\rightarrow\infty$) the CD approach comes very close to the exact result (here represented by MC) whereas the HP overestimates the Curie temperature by 5\%,  RPA underestimates it by 10\% and MFA overestimates it by 40\%.
\begin{figure*}
    \centering
    \includegraphics[scale=0.82]{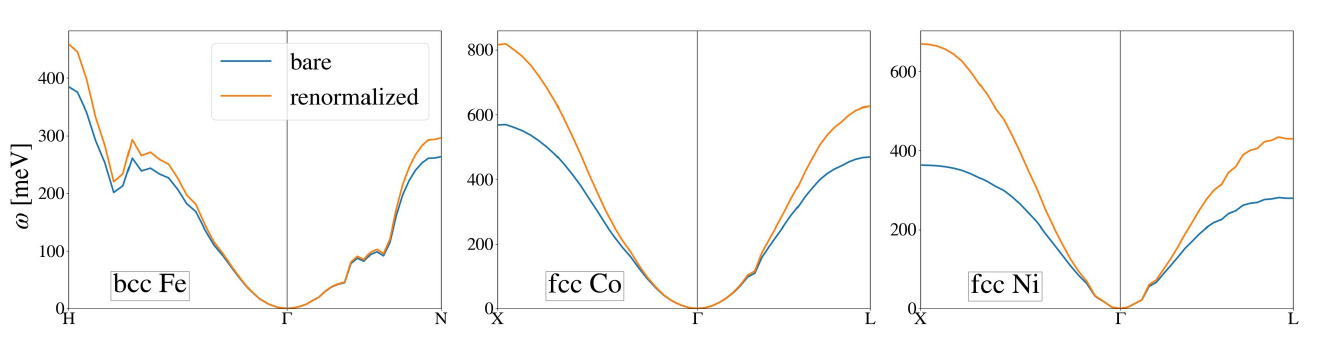}
    \caption{The magnon dispersion of bcc Fe, fcc Co and fcc Ni along high symmetry paths calculated from bare and renormalized MFT.}
    \label{fig:mft}
\end{figure*}

In Ref. \cite{FROBRICH2006223}, QMC calculations for $S$=1/2 revealed that RPA has better agreement with the exact results than CD. In addition, Callen argued in his original paper that the method is probably not accurate for low spin. However, the CD method appears to be the most accurate for the classical limit, where it represents a significant simplification compared to performing classical MC that needs to be carefully converged and may comprise a highly demanding step in terms of computational load. The classical limit is often most appropriate for metals (to the extent that these may be modelled by a Heisenberg model), but not for insulators. In this regard, we note that the ratio between the MFA and RPA Curie temperatures calculated on the basis of the quantum mechanical and classical Heisenberg models respectively is given by $T_\mathrm{c}^\mathrm{QM}/T_\mathrm{c}^\mathrm{Cl}=(S+1)/S$. Commonly applied classical approaches (like MC \cite{Torelli_2019}) are thus expected to underestimate the critical temperature severely - by a factor of three for $S=1/2$ and a factor $5/3$ for $S=3/2$.

\subsection{Curie temperatures of bcc Fe, fcc Ni and fcc Co from first principles}
Here we present the Curie temperatures of bcc Fe, fcc Ni and fcc Co with exchange parameters calculated from DFT. In particular, we apply the magnetic force theorem (MFT) to extract the isotropic parameters $J_{ij}$, which decay slowly as a function of the distance $|\mathbf{R}_j-\mathbf{R}_i|$ due to the itinerant electronic structure \cite{Pajda2001}. 

\subsubsection{MFT calculations}
In the orginal (bare) MFT approach the exchange constants are given by \cite{LIECHTENSTEIN198765,Durhuus2023}
\begin{align}\label{eq:bare_mft}
S^2\tilde J_{ij}=-2\mathrm{Tr}[B_i^\mathrm{xc}\chi^{+-}_\mathrm{KS}B_j^\mathrm{xc}],
\end{align}
where $\chi^{+-}_\mathrm{KS}$ is the static transverse Kohn-Sham susceptibility and $B_i^\mathrm{xc}$ is the exchange-correlation magnetic field truncated in a volume that defines the magnetic site $i$. That is,
\begin{align}
B_i^\mathrm{xc}(\mathbf{r})=B^\mathrm{xc}(\mathbf{r})\Theta(\mathbf{r}\in\Omega_i),
\end{align}
where $\Theta(\mathbf{r}\in\Omega_i)$ is a unit step function that vanishes outside the volume defined by $\Omega_i$. Eq. \eqref{eq:bare_mft} is commonly applied to extract exchange constants and leads to the dispersion 
\begin{equation}
    \tilde{\omega_{\mathbf{q}}} = S(\tilde J_0 - \tilde J_{\mathbf{q}}),
\end{equation}
where $S$ is the site spin, which may be evaluated from the spin density $s^z$ as  $S=\int_{\Omega_i}d\mathbf{r}s^z(\mathbf{r})$. However, it is well established that the expression \eqref{eq:bare_mft} entails a systematic error due to the neglect of constraining fields in the derivation \cite{Bruno2003}. Inclusion of these leads to the renormalized expression
\begin{equation}
    \omega_{\mathbf{q}} = \frac{\tilde\omega_\mathbf{q}}{1 - \tilde\omega_\mathbf{q}/\delta^\mathrm{xc}},
\end{equation}
where $\delta^\mathrm{xc}$ is the site-averaged exchange splitting given by
\begin{equation}
\delta^\mathrm{xc}=-\frac{2}{S}\int_{\Omega_i}d\mathbf{r}s^z(\mathbf{r})B^\mathrm{xc}(\mathbf{r}).
\end{equation}
The renormalization thus becomes important whenever the exchange splitting is small relative to the magnon dispersion.

\begin{table*}[t]
    \centering
    \begin{tabular}{ |p{1.2cm}||p{2cm}|p{2cm}|p{2cm}|p{2cm}|p{1.2cm}|}
         \hline
        & \hspace{0.3in}MFA  & \hspace{0.3in}RPA & \hspace{0.32in}CD & \hspace{0.32in}HP & \hspace{0.15in}Expt \\
         \hline
          bcc Fe &   \hspace{0.03in} 1128 (1004)    &  \hspace{0.06in} 730 \hspace{0.06in}(680)  &  \hspace{0.07in} 827 \hspace{0.05in}(762)  &  \hspace{0.04in} 857 \hspace{0.05in}(760)  &  \hspace{0.09in}1043 \\
        fcc Co &    \hspace{0.06in}1776 (1351)   & \hspace{0.03in}1288 (1062)   & \hspace{0.03in}1412 (1140)   &  \hspace{0.02in}1335 (1010)  & \hspace{0.09in}1388 \\
        fcc Ni &  \hspace{0.12in}602 \hspace{0.02in} (367)  &  \hspace{0.05in} 491 \hspace{0.06in}(328)  &  \hspace{0.05in} 520 \hspace{0.05in}(342) & \hspace{0.08in}450 \hspace{0.06in}(275)   & \hspace{0.15in}627 \\
        \hline
    \end{tabular}
    \caption{The Curie temperatures (in K) of bcc Fe, fcc Co, fcc Ni calculated from renormalized and bare (shown in brackets) MFT using different decoupling approaches. All results are obtained in the classical limit of the methods.}
    \label{table:TcMFT}
\end{table*}

In Fig. \ref{fig:mft} we show the magnon dispersion relations of Fe, Ni and Co obtained from the bare as well as the renormalized expressions. The calculations were carried out with the electronic structure package GPAW \cite{Mortensen2024} using the LDA functional. The \textit{q}-point grids used for the MFT calculations were $\Gamma$-centered and uniform with $N_q^{1/3}=48$ for Fe and Ni and $N_q^{1/3}=36$ for Co. All calculations were done with a plane-wave cutoff of 1000 eV and 9 empty-shell bands. We used site volumes $\Omega_i$ centered on atoms with a radius of half the nearest neighbor distance. This yields site spins of 1.117, 0.3249, 0.8345 (in units of $\hbar$) and average exchange splittings $\delta^\mathrm{xc}$ of 2.3767, 0.79239 and 1.8769 eV for Fe, Ni and Co respectively. The site moments may be compared with the total spin per unit cell, which is 1.10, 0.314 and 0.815 respectively. The difference between the local and unit cell spin may be regarded as an inherent uncertainty in the method, which crucially relies on mapping an itinerant system to a localized spin model. The calculated magnon dispersions exhibit a significant effect of the renormalization in Ni and Co due to the rather small exchange splitting in these compounds, whereas the renormalization is less pronounced in Fe. The renormalized MFT dispersions are in excellent agreement with magnonenergies obtained from time-dependent density functional theory (TDDFT) \cite{Skovhus2021}, whereas the bare MFT expressions underestimate the TDDFT results. The actual magnetic response of these systems is prone to rather strong Landau damping---in particular when approaching the Brillouin zone boundary---and the definition of a magnon energy from the TDDFT spectral function becomes ambiguous in some parts of the Brillouin zone \cite{Skovhus2021}. The MFT approach will, however, by construction yield a magnon energy at any $\mathbf{q}$ and the noisy features of the magnon dispersions of Fig. \ref{fig:mft} are related to nontrivial (not just Lorentzian broadening) interactions with the Stoner continuum. Nevertheless, we may use the MFT magnon energies to evaluate critical temperatures based on the different decoupling schemes discussed above, keeping in mind that the Heisenberg modelling of such systems is rather severely limited by the itinerant nature of the compounds.

\subsubsection{Critical Temperatures}
The Curie temperature of bcc Fe, fcc Co and fcc Ni obtained from MFA, RPA, CD and HP are tabulated in Tab. \ref{table:TcMFT}. 
As these compounds are metallic, the spin in site-based models are best regarded as continuous variables and the calculations were thus performed in the classical limit. In order to due this for the HP and Callen methods, we calculate $T_\mathrm{c}(S)/S^2$ in the limit of $S\rightarrow\infty$ and then obtain the critical value by multiplying by the actual site-spin obtained from first principles. In the case of RPA, the classical limit can be obtained from Eq. \eqref{Tc3d-rpa} by replacing $S(S+1)$ by $S^2$. We state the numbers obtained from bare MFT (in brackets) as well as renormalized MFT. Comparing the TDDFT calculations of these materials \cite{Skovhus2021} shows that the magnetic excitation spectrum of fcc Co is by far the simplest in the sense that the spectral function is well approximated by a Lorentzian line-shape centered on a value, which is in excellent agreement with the prediction from renormalized MFT. We thus regard the exchange parameters and magnon dispersion of fcc Co as the most reliable. It is observed that the Curie temperature of fcc Co is in agreement with the experimental value if one uses MFA combined with {\it bare} MFT. This is a good example of error cancellation that could lead one to believe that the combination of these two approaches provide a good account of the magnetic properties. However, from Fig. \ref{fig:s-classical3d} it is clear that MFA is expected to overestimate the Curie temperature rather severely. Indeed, when applying the more accurate renormalized MFT dispersion, MFA leads to an overestimation of the critical temperature by 28 \%, which is significant although less severe than for the the simple cubic lattice due to the higher coordination number in the fcc lattice. On the other hand, the RPA, CD and HP methods predict Curie temperatures rather close to the experimental one when based on the renormalized MFT, whereas they underestimate it when combined with bare MFT (due to underestimated magnon energies). 

\begin{figure*}[tb]
     \centering
     \includegraphics[scale=0.82]{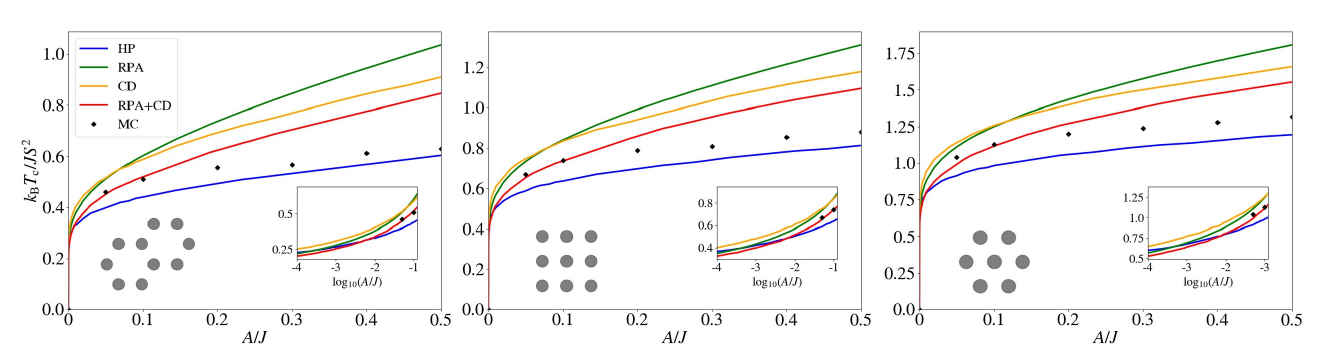}
     \caption{Calculated critical temperature as a function of single-ion anisotropy for honeycomb, square and hexagonal lattices in the classical limit.}
     \label{fig:sia-classical}
\end{figure*}
The situation is somewhat more complicated for Fe and Ni, where the renormalized MFT results do not provide as accurate an account of the magnetic excitations (compared to TDDFT) in the entire Brillouin zone. For Fe, the MFA, based on renormalized MFT, overestimates the Curie temperature whereas RPA, HP and CD underestimates it. The failure of the latter methods is likely related to fact that the transverse susceptibility of Fe extends to high energies (up to 1 eV) in large regions of the Brillouin zone (not captured by MFT) due to significant mode branching and deviation from Lorentzian lineshapes in these regions. Again, we observe that the error is smaller in the HP and CD methods compared to RPA, in agreement with the cubic model of Sec. \ref{sec:cubic}. The closest agreement with the experimental value is obtained with the MFA based on bare MFT energies, but again relies on error cancellation between underestimated exchange parameters and overestimated Curie temperatures in the MFA. For fcc Ni the situation is similar except that the HP method seems to have the worst performance. The fact that DFT based methods (combined with RPA) underestimate the Curie temperature has been discussed at some length in the literature \cite{Bruno2003, Katsnelson2004} and has been used to argue for the importance of renormalization. However, the experimental dispersion relation for Ni appears to be in better agreement with bare MFT \cite{Katsnelson2004}. Nevertheless, it is clear that the renormalized MFT dispersion is much closer to results obtained from TDDFT (in the adiabatic local density approximation) \cite{Skovhus2021} and the overestimated magnon dispersion in renormalized MFT is likely because LDA itself fails to provide an accurate representation of the ground state \cite{Aryasetiawan1992,Muller2019}. 

\subsection{2D model lattices}
We now turn to 2D systems where magnetic anisotropy plays a crucial role for magnetic order at finite temperature and an accurate decoupling of the single-ion anisotropy therefore becomes decisive. In Ref. \cite{FROBRICH2006223} it was shown that the CD scheme compares well with an exact treatment if the anisotropy constant is not too large relative to the exchange constants, which is typically satisfied in real materials. On the other hand, it was also shown that for low spin the CD performs poorly compared to RPA in isotropic systems, leading to a suggestion of mixing the schemes (RPA+CD), such that all terms except the single-ion anisotropy are decoupled according to \eqref{rpa-approx} and the corrections from single-ion anisotropy are included through the Callen decoupling Eq. \eqref{callen_correction}. 

In Fig. \ref{fig:sia-classical} we show the critical temperatures obtained in the classical limit for hexagonal, square and honeycomb lattices with nearest neighbor interaction $J$. We refer to Appendix \ref{sec:multisite} for a generalization of the methods to an arbitrary number of magnetic sites in the unit cell (we need this for the two sites of the honeycomb lattice). We display the results calculated with HP, RPA, RPA+CD as well as the classical MC results.
Even though classical MC may serve as the exact reference, simulations in the limit of $A/J \rightarrow 0$e difficult to converge and less reliable because the $T_\mathrm{c}$ predicted by heat capacity and magnetization starts to diverge. So, we limited ourselves to a range where classical MC is reliable. In all cases we observe the HP method to underestimate critical temperature although it acquires the best agreement with MC when $A/J$ becomes sufficiently large. In addition, RPA yields the largest deviation for large anisotropies, while the CD approach yields the worst results for small anisotropies. The RPA+CD approach performs better than both the RPA and CD for any value of the anisotropy, but performs significantly worse than the HP method for large anisotropies. It has previously been shown that all the decoupling schemes fail rather dramatically in the limit of $A/J\rightarrow\infty$. In that case the transverse fluctuations are frozen out and the Heisenberg Hamiltonian reduces to the Ising
model, but all of the decoupling schemes yields $T_\mathrm{c}\propto A$ for $A/J\rightarrow\infty$ \cite{Torelli_2019,FROBRICH2006223} rather than $T_\mathrm{c}\rightarrow T_\mathrm{c}^\mathrm{Ising}$. The origin of this behaviour can be exemplified by the MFA where the mean-field decoupling of the single-ion anisotropy yields $3k_\mathrm{B}T_\mathrm{c}=S(S+1)(J_0+2A)$, while the exact treatment must become independent of $A$ in the limit of $A/J_0\rightarrow\infty$. Here we just note that the Curie temperatures obtained from RPA exhibit the fastest increase with respect to $A$ and breaks the Ising limit rather quickly, whereas the HP method has the slowest increase in Curie temperature and only breaks the Ising limit at a rather rather large value of $A/J$. Although we have only compared with exact results in the classical limit, these calculations could indicate that the RPA+CD method has a superior performance for 2D systems with single-ion anisotropy in the regime where $A/J < 0.2$.

\subsection{Application to monolayer CrI$_3$}
The presence of ferromagnetic order in 2D was first demonstrated for the case of monolayer CrI$_3$ \cite{Huang2017}. Subsequently, there has been a wide variety of attempts to reproduce the Curie temperature based on first principles predictions of exchange constants combined with thermodynamics based on the Heisenberg model. The magnetic Cr atoms form a honeycomb lattice with $S=3/2$, which constitutes the basic ingredient in site-based magnetic modelling. Thermodynamic approaches span a wide variety of methods from Ising modelling over classical MC, HP and RPA based on the Heisenberg model. However, the predicted values of exchange constants vary by a factor of three \cite{Olsen2019} depending on the applied method and to our knowledge there has not been attempts to critically assess the performance of the subsequent thermodynamic modelling. As will be shown below, the predicted Curie temperature may vary by a factor of two depending on the thermodynamic method for a given set of exchange parameters.
\begin{figure*}[ht!]
     \centering
     \includegraphics[scale=0.84]{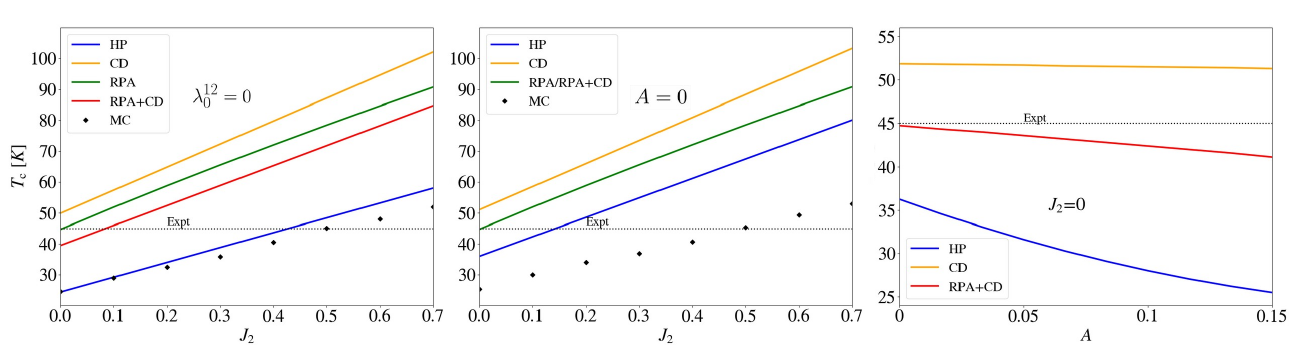}
     \caption{Curie temperatures of CrI$_3$ obtained with nearest neighbor interactions $J_1=2.0$ meV and anisotropy parameters consistent with the experimental spin wave gap of $\Delta=0.3$ meV. Left: Curie temperature as function of $J_2$ with $\lambda^{12}_0=0$ and $A$ determined from Eqs. \eqref{delta_rpa}-\eqref{delta_cd}. Center: Curie temperature as function of $J_2$ with $A=0$ and $\lambda^{12}_0$ determined from Eqs. \eqref{delta_rpa}-\eqref{delta_cd}. Right: Curie temperature as a function of the single-ion anisotropy $A$ with $J_2=0$ and $\lambda^{12}_0$ determined from Eq. \eqref{delta_cd}.}
    \label{fig:cri3}
\end{figure*}

We will not delve into different methodologies for obtaining exchange constants but only focus on the different approaches for calculating the Curie temperature given a set of Heisenberg parameters. Although these are not known exactly, the spinwave gap and the optical magnon at the $\Gamma$-point for bilayer CrI$_3$ have been determined experimentally to be 0.3 meV and 18.4 meV respectively \cite{Cenker2021} and these values puts strong bounds on the exchange constants. In particular, within the model defined by Eqs. \eqref{heisenbrghamiltonian} and \eqref{2dheisenberghamiltonian}, the spinwave gap at $T=0$ obtained in either semi-classical spinwave theory (relevant for classical MC) or RPA is given by
\begin{equation}\label{delta_rpa}
    \Delta^\mathrm{RPA/MC} = 2AS + S\lambda^{12}_0
\end{equation}
while in CD and HP it is given by
\begin{equation}\label{delta_cd}
    \Delta^\mathrm{CD/HP} = A(2S-1) + S\lambda^{12}_0.
\end{equation}
In addition, within RPA, CD and HP the optical magnon at $\mathbf{q=0}$ is situated at
\begin{equation}\label{optical}
    \omega^\mathrm{opt}_{\mathbf{q}=0} = 2S J^{12}_0 + \Delta,
\end{equation}
where $J^{12}_0$ and $\lambda^{12}_0$ are the sum of inter-sublattice exchange constants and anisotropic exchange constants respectively (defined in Eqs.\eqref{2siteJab} and \eqref{2SiteAE}). We have ignored the effect of of intra-sublattice anisotropic exchange ($\lambda^{11}_0$) in our model. Most first principles methods agree that the exchange constants vanish beyond third nearest neighbors and the isotropic model is thus defined by $J_1$, $J_2$ and $J_3$, which yields $J^{12}_0=N_\mathrm{nn}(J_1+J_3)$ with $N_\mathrm{nn}=3$. The third nearest neighbor exchange is typically predicted to be rather small and we may thus use Eq. \eqref{optical} to get $J_1=2.0$ meV, while $J_2$ remains undetermined by experiements. Similarly, depending on whether one uses RPA/MC or HP/CD for modelling, the single-ion anisotropy and sum of anisotropic exchange constants may be obtained from Eq. \eqref{delta_rpa} or Eq. \eqref{delta_cd} respectively. Now, it is not clear whether the spin-wave gap in CrI$_3$ originates from either of these constants. Specifically, LDA and PBE predict equal contribution from $A$ and $\lambda^{12}_0$ while LDA+U and PBE+U predict that $A\sim0$ and the spin-wave gap thus mainly originate from anisotropic exchange \cite{Olsen2019}. However, all of these functionals predict a spin-wave gap of $\sim0.7-1.0$ meV, which is more than a factor of two larger than the experimental gap and the reliability of any of these predictions may be questionable.

In Fig. \ref{fig:cri3} we show the dependence of the critical temperatures obtained with classical MC, HP, RPA, CD and RPA+CD, while the parameters $J_2$, $\lambda^{12}_0$ and $A$ are varied. In all cases we have taken $J_1=2.0$ meV, $J_3=0$ (consistent with the optical magnon at $\Gamma$) and with the spin wave gap fixed to the experimental value of 0.3 meV. In the first two plots, we show the results for the two limiting cases of $\lambda^{12}_0=0$ and $A=0$ where the remaining anisotropy parameter is obtained from Eqs. \eqref{delta_rpa} or \eqref{delta_cd}. We also show the variation when the anisotropy (represented by the experimental spin wave gap) is mapped to different values of $A$(and $\lambda^{12}_0)$, while enforcing $\Delta=0.3$ meV and $J_2=0$. We first note that the Curie temperature in the classical MC, RPA (not shown, but invariant by construction) and CD approaches are largely independent on whether the spin wave gap originates from single-ion anisotropy or anisotropic exchange. The RPA+CD exhibits moderate variation, while the HP method has strong dependence on the origin of anisotropy. We believe that this represents a severe problem with the description of single-ion anisotropy in the HP method. In particular, one may show that the Curie temperature in the HP method will decrease with increasing anisotropic exchange (and increasing spin wave gap) for a fixed value of $A$, which seems unreasonable. The thermal renormalization of the spin-wave gap is included through the term $-4A\phi$ in Eq. \eqref{hp_correction}, which leads to a (too) rapid closing of the gap at elevated temperatures. In contrast, the renormalization is rescaled by $\langle S^z\rangle^2$ in the CD method, which leads to a slower and continuous closing of the spinwave gap when the temperature is increased.

Based on the classical limit illustrated in Fig. \ref{fig:sia-classical} as well as previous simulations on spin-1/2 systems, it appears likely that the RPA+CD method comprises the most accurate approach. Based on the present results it is, however, not possible to assert this statement, since the exact value of $J_2$ is not known. LDA calculations yield $J_1+J_3=2.0$ meV and $J_2=0.6$ meV, while more accurate calculations based on many-body perturbation theory \cite{Olsen2021} as well as experiments on bulk CrI$_3$ \cite{Chen2018a,Lee2020} imply that $J_2$ should be on the order of 0--0.2 meV. However, the magnetic excitation spectrum contains a significant gap at the $K$-point as well \cite{Chen2018a, Lee2020}, which cannot be accounted for by the present model \cite{Olsen2021,Delugas2023}. Nonetheless, we wish to emphasize the uncertainty in obtaining critical temperatures for 2D materials from simple decoupling schemes such as the ones studied here, where the predicted Curie temperature can differ by more than a factor of two depending on the method of choice. It may therefore be unreliable to benchmark first principles simulations against experimental Curie temperatures of 2D materials obtained by the methods discussed here. The worst approach for non-classical systems is arguably the classical MC, which severely underestimates the Curie temperature. This can be understood from the MFA in 3D, which will predict 60 \% of the quantum value for a spin-3/2 system and it is highly likely that such errors are inherited by a classical treatment of 2D systems. In addition, the HP method appears to become unreliable when single-ion anisotropy is included.

\section{Conclusion and outlook}\label{sec:conclusion}
We have compared the performance of classical MC, MFA, RPA, HP and the CD decoupling schemes for calculating Curie temperatures of ferromagnets. In the classical limit, the MC simulations are regarded as the exact reference and for a cubic lattice with nearest neighbor isotropic interactions, the CD method becomes highly accurate while the HP overestimates the Curie temperatures slightly and RPA underestimates it. In contrast, the MFA yields a significant overestimation (39 {\%}) of the Curie temperatures in the classical limit. It has, however, previously been shown that the RPA method is more accurate than the CD for $S=1/2$ systems \cite{FROBRICH2006223}. On the other hand, from Fig. \ref{fig:s-classical3d} it is clear that the HP decoupling interpolates between RPA and the CD results when $S$ is increased from $1/2$ to the classical limit and the HP method could perhaps be regarded as the best comprise for isotropic systems with arbitrary spin.
\begin{table}[tb]
\begin{center}
\begin{tabular}{| l | c |}
\hline
 3D Metals & CD / HP / Classical MC \\ 
 \hline
 3D Insulators & RPA  / HP\\
 \hline
 2D Insulators - SIA  & RPA+CD \\ 
\hline
 2D Insulators - no SIA & HP \\ 
\hline
\end{tabular}
\end{center}\label{tab:recommended}
\caption{Recommended approaches for extracting Curie temperatures in 3D metals/insulators and 2D materials. The RPA+CD method is equivalent to RPA except for the single-ion anisotropy terms, which are treated according to the Callen decoupling method. For 2D systems the best approach depends on whether or not the method includes single-ion anisotropy (SIA).}
\end{table}

The methods were then applied to the itinerant ferromagnets bcc Fe, fcc Co and fcc Ni with exchange constants obtained from the (renormalized) magnetic force theorem. In these cases the classical limit seems to be well justified and as expected the CD yields the overall best agreement with experimental Curie temperatures. Nevertheless, in the cases of Fe and Ni, the predicted Curie temperatures are still significantly lower than experimental values, which is likely due to the drastic oversimplification involved when mapping the spectral function of itinerant magnets into a site-based model, which does not take Landau damping into account. To this end, we note that the best results are obtained in the case of fcc Co, where TDDFT calculations have previously shown the best agreement with a single (broadened) magnon dispersion \cite{Skovhus2021}. It would be highly interesting to generalize the present approach such that it is based on the spectral function $A^{+-}(\mathbf{q},\omega)$ obtained from TDDFT. A simple generalization would be to replace the Bose factors according to \cite{Paischer2021}
\begin{align}
n^\mathrm{B}_\mathbf{q}(\omega_\mathbf{q})\rightarrow\int_0^\infty d\omega\frac{A^{+-}(\mathbf{q},\omega)}{e^{\hbar\omega/k_\mathrm{B}T}-1}.
\end{align}
Such an approach would take the lineshape of the magnon modes as well as possible mode branching into account, but would comprise a tremendous computational task due to the requirement of densely sampled TDDFT calculations in the entire Brillouin zone. 

For the case of 2D materials, the conclusions become a bit more more blurry. The reason is that magnetic anisotropy plays a crucial role for magnetic order and the inclusion of such terms in the Heisenberg model makes the decoupling schemes more dubious. By default, RPA fails to reproduce the lack of magnetic order for $S=1/2$ systems with single-ion anisotropy, while the HP method appears to severely overestimate the thermal renormalization of the spin-wave gap and the CD method overestimates Curie temperatures when the spin-wave gap becomes small. We have exemplified these problems in the classical limit for honeycomb, square and hexagonal lattices where we again compare to classical MC calculations. In all cases, the RPA+CD (RPA method for isotropic and CD for anisotropic terms) yields good agreement with classical MC for low values of anisotropy ($A/J<0.2$). For large anisotropies, all the methods break down and asymptotically approach a linear relation between the single-ion anisotropy and the critical temperature, whereas the exact result should approach the Ising limit asymptotically. On the other hand, in the quantum regime the classical MC leads to significant underestimation of critical temperature and therefore does not comprise a viable alternative to the quantum mechanical decoupling schemes.

We applied the methods for obtaining Curie temperatures in 2D to the case of CrI$_3$, where we focused on Heisenberg parameters extracted from experiments. This constitutes a rather incomplete description of the system and we left the next nearest exchange coupling as a free parameter. It is, however, evident that the predicted Curie temperatures are rather sensitive to the applied method. It is difficult to judge which of the Green's function methods (RPA, CD and RPA+CD) that is most appropriate for CrI$_3$, but based on the previous calculations we believe that RPA+CD should be the preferred approach. The HP method, yields significantly lower Curie temperatures (comparable to classical MC) 
if the magnetic anisotropy is governed by single-ion anisotropy, whereas it seems to provide the best agreement with experiments if the anisotropy is governed by anisotropic exchange. We believe that the HP probably constitutes the most accurate method if single-ion anisotropy can be neglected, but it should be avoided if single-ion anisotropy plays a prominent role. 
For classical MC, the Curie temperature is expected to be underestimated in general. As a result, previous reports on theoretical critical temperature that are in agreement with experimental ones, are likely prone to a fortuitous error cancellation (due to overestimated exchange constants) whenever these are based on classical MC combined with DFT \cite{Torelli_2019,Torelli2020,Kabiraj2022,Wines2023}. 

In this way, the accuracy of different decoupling schemes in 3D is not directly transferable to 2D magnets due to the inherent problems associated with decoupling the single-ion anisotropy. While there are plenty of bulk 3D materials where Heisenberg parameters have been determined rigorously from inelastic neutron scattering, the parameters for 2D magnets are more scarce. The acquisition of such data is crucial for properly benchmarking the performance of different decoupling schemes in 2D and the experimental extraction of accurate parameters thus seems pertinent for computational progress in the field. Nevertheless, based on the results in the present paper as well as previous works we believe that one may formulate recommendations that depend on the type of system under scrutiny. In Tab. \ref{tab:recommended}, we summarize the methodologies that are expected to be most accurate for different types of materials and hope that this may serve as useful guidelines for future calculations.

\section{Acknowledgements}
The authors acknowledge support from the Villum foundation Grant No. 00029378.

\appendix
\section{Decoupling methods for multiple magnetic sites in the unit cell}\label{sec:multisite}
Here we state the generalizations of the renormalized magnon dispersion obtained in the various decoupling schemes for systems with multiple sites in the unit cell. The Hamiltonian of interest is thus
\begin{equation}
    H = -\frac{1}{2}\sum_{aibj} J_{ij}^{ab}\mathbf{S}_{ai} \cdot \mathbf{S}_{bj} -\frac{1}{2}\sum_{aibj} \lambda_{ij}^{ab}S_{ai}^z  S_{bj}^z  - A \sum_{ai}(S_{ai}^z)^2,
\end{equation}
where $\mathbf{S}_{ai}$ is the spin operator for site $a$ in unit cell $i$ and we assume $J_{00}^{aa}=\lambda_{00}^{aa}=0$.

\subsection{Holstein-Primakoff bosonization}
For the HP method, one introduces a bosonic transformation for each site in the unit cell:
\begin{align}
    S_{ai}^+&=\sqrt{2S_a-a_{ai}^\dag a_{ai}}a_{ai}, \\
    S_{ai}^-& =a_{ai}^\dagger\sqrt{2S_a-a_{ai}^\dagger a_{ai}}, \\
    S^z_{ai}& =S_a-a_{ai}^\dag a_{ai}.
\end{align}
For simplicity we restrict ourselves to the case of a ferromagnetic ground state  $|0\rangle$  with equivalent atoms (having the same spin $S_a=S$), but the method is straightforward to generalize to multiple sites with different values of $S_a$. The thermal decoupling of fourth order terms then yields
\begin{align}
H^\mathrm{HP}=E_0+\sum_{ab\mathbf{q}}a^\dag_{a\mathbf{q}}H_{\mathbf{q}}^{ab}a_{b\mathbf{q}},
\end{align}
where $E_0$ is the ground state energy (including thermal corrections) and
\begin{align}
    H^{ab}_\mathbf{q} =& S\delta^{ab}\sum_{c}(J^{ac}_0 +\lambda^{ac}_0 )- SJ^{ab}_\mathbf{q}+A(2S-1)\delta^{ab}\notag\\
    +& \frac{1}{N_q} \sum_\mathbf{q^\prime}\bigg\{ J^{ab}_\mathbf{q} \frac{n^{aa}_\mathbf{q^\prime}+ n^{bb}_\mathbf{q^\prime}}{2} -(J^{ab}_{\mathbf{q}-\mathbf{q}'}+\lambda^{ab}_{\mathbf{q}-\mathbf{q}'} ) n^{ba}_\mathbf{q^\prime} \notag\\
    +&\delta^{ab}\sum_c\Big[\frac{J^{ac}_\mathbf{q^\prime} n^{ac}_\mathbf{q^\prime}+J^{ca}_\mathbf{q^\prime} n^{ca}_\mathbf{q^\prime} }{2}
    -(J^{ac}_0 + \lambda^{ac}_0 ) n^{cc}_\mathbf{q^\prime}\Big] \notag\\
    -&\delta^{ab}4An^{aa}_\mathbf{q^\prime}\bigg\}.
\end{align}
where $n^{ab}_\mathbf{q}=\langle a^\dag_{\mathbf{q}a}a_{\mathbf{q}b}\rangle$ and 
\begin{align}
    J^{ab}_\mathbf{q} &= \sum_i J^{ab}_{0i} e^{i \mathbf{q} \cdot \mathbf{R}_i} ,\label{2siteJab}\\
    \lambda^{ab}_\mathbf{q} &=q \sum_i \lambda^{ab}_{0i} e^{i \mathbf{q} \cdot \mathbf{R}_i}. \label{2SiteAE}
\end{align}
Diagonalizing this Hamiltonian leads to a set of eigenvalues $\omega_{\mathbf{q}n}$ and eigenstates $a^\dag_{\mathbf{q}n}|0\rangle$ with $a^\dag_{\mathbf{q}n}=\sum_av_{\mathbf{q}na}a^\dag_{\mathbf{q}a}$. We can thus write
\begin{align}
    n^{ab}_\mathbf{q}=\sum_{mn}v_{\mathbf{q}ma}^*v_{\mathbf{q}nb}\langle a^\dag_{\mathbf{q}m}a_{\mathbf{q}n}\rangle=\sum_{n}v_{\mathbf{q}na}^*v_{\mathbf{q}nb}n^\mathrm{B}(\omega_{\mathbf{q}n}),
\end{align}
where $n^\mathrm{B}(\omega_{\mathbf{q}n})$ is the Bose distribution for magnon branch $n$. 
Finally, the thermal average of the spin at a particular site is given by
\begin{align}
    \langle S^z\rangle=S-\phi,
\end{align}
where $\phi$ is now given by
\begin{align}\label{eq:phi_multiple}
    \phi=\frac{1}{N_qN_a}\sum_{\mathbf{q}n}n^\mathrm{B}(\omega_{\mathbf{q}n}),
\end{align}
and $N_a$ is the number of magnetic sites in the unit cell.

\subsection{Random Phase Approximation}
With multiple sites in the unit cell we consider the retarded Green's function
\begin{equation}
    G^{+-}_{aibj}(t) = \langle \langle S_{ai}^+(t) ; S_{bj}^-\rangle \rangle.
\end{equation}
The equation of motion becomes
\begin{align}
    \omega G^{+-}_{aibj}(\omega) = 2 \langle S_a^z \rangle \delta_{ij} \delta_{ab} + \langle \langle [S_{ai}^+,H] ; S_{bj}^- \rangle \rangle_\omega,
\end{align}
where the commutator yields
\begin{align}
    [S_{ai}^+,H] &= \sum_{ck} J_{aick} ( S_{ai}^+ S_{ck}^z  - S_{ai}^zS_{ck}^+ ) \notag\\ &+ \sum_{ck} \lambda_{aick} S_{ai}^+ S_{ck}^z \notag \\& + A(S_{ai}^z S_{ai}^+ -  S_{ai}^+S_{ai}^z).
\end{align}
One then makes the RPA approximation
\begin{equation}
     \langle \langle S_{ai}^+ S_{ck}^z  ; S_{bj}^-  \rangle \rangle \approx \langle S^z_c \rangle \langle \langle S_{ai}^+  ; S_{bj}^- \rangle \rangle,
\end{equation}
which gives
\begin{align}
    \omega G^{+-}_{\mathbf{q}ab}(\omega) &=2 \langle S_a^z \rangle \delta_{ab} +  G^{+-}_{\mathbf{q}ab}(\omega)\sum_c \langle S_c^z \rangle (J^{ac}_0 + \lambda^{ac}_0 )  \notag\\
    &- \langle S_a^z\rangle \sum_c J^{ac}_\mathbf{q} G^{+-}_{\mathbf{q}cb}(\omega) + 2A \langle S_a^z \rangle G^{+-}_{\mathbf{q}ab}(\omega).
\end{align}
This can be written as
\begin{align}\label{eqnGfn2site}
    \sum_c(\omega\delta_{ac} -H_{\mathbf{q}}^{ac})G^{+-}_{\mathbf{q}cb}(\omega) =2 \langle S_a^z \rangle \delta_{ab},
\end{align}
with
\begin{align}\label{multisite-H_rpa}    H_{\mathbf{q}}^{ab}=\delta_{ab}\sum_c \langle S_c^z \rangle (J^{ac}_0 + \lambda^{ac}_0 ) + \delta_{ab}2A \langle S_a^z \rangle - \langle S_a^z\rangle J^{ab}_\mathbf{q}.
\end{align}
The Green's function is thus be obtained as
\begin{align}\label{multi-site-Gfn}
    G^{+-}_{\mathbf{q}ab}(\omega) =2\langle S_b^z \rangle\sum_n \frac{U_{\mathbf{q}an}U^*_{\mathbf{q}bn}}{\omega-\omega_{\mathbf{q}n}+i\eta},
\end{align}
where $\omega_{\mathbf{q}n}$ are the eigenvalues of $H_{\mathbf{q}}^{ab}$, which is diagonalized by the matrix $U_{\mathbf{q}an}$. Finally, for equivalent magnetic sites, the magnetization is given by an expression that is formally equivalent to Eq. \eqref{mag}, but where $\phi$ is now given by the multi-site generalization \eqref{eq:phi_multiple}.

\subsection{Callen Decoupling}
For the Callen decoupling scheme the higher order Green's function is approximated by
\begin{align}
    \langle \langle S_{ck}^z S_{ai}^+; S_{bj}^- \rangle \rangle &\approx \langle S^z \rangle \langle \langle S_{ai}^+; S_{bj}^- \rangle \rangle - \alpha \langle S_{ck}^- S_{ai}^+ \rangle \langle \langle S_{ck}^+; S_{bj}^- \rangle \rangle.
\end{align}
We again limit ourselves to magnetic systems with equivalent spins($\langle S_a^z \rangle = \langle S_b^z \rangle = \langle S^z \rangle$) to avoid the ambiguity related to the spin dependence of $\alpha$ and thus use $\alpha = \frac{\langle S^z \rangle}{2S^2}$. Defining $\psi_{aibj}=\langle S^-_{bj}S^+_{ai}\rangle$ this gives 
\begin{align}
    \omega G^{+-}_{\mathbf{q}ab}(\omega) &= 2 \langle S^z \rangle \delta_{ab} +  G^{+-}_{\mathbf{q}ab}(\omega)\langle S^z \rangle\sum_c   (J^{ac}_0 + \lambda^{ac}_0 )  \nonumber \\
    &-\langle S^z \rangle\sum_cJ^{ac}_\mathbf{q}G^{+-}_{\mathbf{q}cb}(\omega) \nonumber \\ 
    &+ \frac{\langle S^z \rangle}{2S^2N_q}\sum_{c\mathbf{q^\prime}} \Bigl\{ J^{ac}_\mathbf{q^\prime} \psi^{ca}_\mathbf{q^\prime} G^{+-}_{\mathbf{q}ab}(\omega)  \nonumber \\&- (\lambda^{ac}_\mathbf{q-q^\prime} + J^{ac}_\mathbf{q-q^\prime}) \psi^{ac}_\mathbf{q^\prime} G^{+-}_{\mathbf{q}cb}(\omega) \Bigr\}\nonumber \\&+ A\Bigl(2\langle S^z \rangle - \frac{\langle S^z \rangle}{S^2}( 1 + \frac{1}{N_q}\sum_\mathbf{q^\prime} \psi^{aa}_\mathbf{q^\prime} ) \Bigr)G^{+-}_{\mathbf{q}ab}(\omega). \label{twositeCD}
\end{align}
One may now proceed as in the case of RPA but with
\begin{align}\label{multisite-H_cd}
    H^{ab}_\mathbf{q} &= \delta_{ab} \langle S^z \rangle \sum_c(J^{ac}_0 + \lambda^{ac}_0 ) + \langle S^z \rangle J^{ab}_\mathbf{q}  \nonumber \\ &+ \frac{\langle S^z \rangle}{2S^2N_q}\sum_{\mathbf{q^\prime}} \Bigl\{ \delta_{ab} \sum_c J^{ac}_\mathbf{q^\prime} \psi^{ca}_\mathbf{q^\prime}  - (\lambda^{ab}_\mathbf{q-q^\prime} + J^{ab}_\mathbf{q-q^\prime}) \psi^{ac}_\mathbf{q^\prime} \Bigr\}\nonumber \\&+ \delta_{ab} A\Bigl(2\langle S^z \rangle - \frac{\langle S^z \rangle}{S^2}( 1 + \frac{1}{N_q}\sum_\mathbf{q^\prime} \psi^{aa}_\mathbf{q^\prime} ) \Bigr),
\end{align}
and $\psi^{ab}_\mathbf{q}$ may be obtained from the fluctuation-dissipation theorem as
\begin{equation}
 \psi^{ab}_\mathbf{q} = 2\langle S^z \rangle \sum_n U_{\mathbf{q}an}U^*_{\mathbf{q}bn} n^\mathrm{B}(\omega_{\mathbf{q}n}).
\end{equation}
The Green's function is again given by Eq. \eqref{multi-site-Gfn} but with $\omega_{\mathbf{q}n}$ being the eigenvalues of Eq. \eqref{multisite-H_cd}, which are diagonalized by the corresponding $U_{\mathbf{q}an}$.

\bibliographystyle{unsrt}
\bibliography{references}

\end{document}